# Mott insulator tuning via structural distortion in monolayer 1T-NbSe$_2$

Zhen-Yu Liu[1,&], Shuang Qiao[2,&], Qiao-Yin Tang[1], Zi-Heng Ling[1], Wen-Hao Zhang[1], Hui-Nan Xia[1], Xin Liao[1], Wen-Hao Mao[1], Jing-Tao Lü[1], Bing Huang[2], and Ying-Shuang Fu[1,*]

1. School of Physics and Wuhan National High Magnetic Field Center, Huazhong University of Science and Technology, Wuhan 430074, China

2. Beijing Computational Science Research Center, Beijing 100093, China

& These authors contribute equally to this work.
*yfu@hust.edu.cn

**Mott state in 1T-TaS$_2$ is predicted to host quantum spin liquids (QSL). However, its insulating mechanism is controversial due to complications from interlayer coupling. Here, we study the Mott state in monolayer 1T-NbSe$_2$, an electronic analogy to TaS$_2$ exempt from interlayer coupling, using spectroscopic imaging scanning tunneling microscopy and first principles calculations. Monolayer NbSe$_2$ surprisingly displays two types of Star-of-David (SD) motifs with different Mott gap sizes, that are interconvertible via temperature variation. And, bilayer 1T-NbSe$_2$ shows Mott collapse by interlayer coupling. Our calculation unveils the two types of SDs possess distinct structural distortions, altering the effective Coulomb energies of the central Nb orbital. Our calculation suggests the Mott gap, the same parameter for determining the QSL regime, is tunable with strain. This finding offers a general strategy for manipulating the Mott state in 1T-NbSe$_2$ and related systems via structural distortions, which may be tuned into the potential QSL regime.**

Mott insulator arises from localization of electrons when Coulomb energy overwhelms the kinetic energy, i.e. the single particle bandwidth, which opens a gap to the otherwise metallic system [1[3]. In two-dimensional (2D) Mott insulators with triangular lattice and antiferromagnetic coupling, the inclusion of ring exchange terms renders the system feasible as a candidate for hosting quantum spin liquid (QSL) [4[7], a state of entangled spins without magnetic order [8-10]. One notable Mott system of such kind is 1T-TaS$_2$ [11[][14], a layered van der Waals crystal. It experiences a charge density wave (CDW) transition at low temperatures, resulting in an ordered lattice of Star-of-David (SD) motifs [15]. In each SD, there is an unpaired electron in the center, which possess prominent on-site Coulomb repulsion and reduced hopping integral, due to the large spatial spacing among neighboring centers [16[18]. While this scenario has been widely accepted, the interlayer coupling is largely neglected. Recently, increasing studies have recognized the indispensable role of interlayer coupling, challenging the ground state of the system with conventional band insulating mechanisms [19-[22]. This raises the necessity of studying single-layer 1T-TaS$_2$ to clarify these debates. However, a different CDW order exists in the exfoliated TaS$_2$ single-layer [23], and the sample quality of grown monolayer films with molecular beam epitaxy (MBE) achieved until very recently is not compatible to that of the bulk [24].

To overcome that hurdle, an alternative arena is to study related compounds as exemplified in 1T-TaSe$_2$ and 1T-NbSe$_2$, which share similar Mott physics as that in TaS$_2$. Such monolayer compounds have been successfully synthesized with MBE, and display Mott insulating states with the SD lattices [25[26]. These alternative systems not only render it possible to study monolayer Mott systems that are devoid of interlayer coupling, but also are of importance of their own light, due to their different strength of electron-phonon coupling, Coulomb energy, *etc*. And, the Mott states are intimately connected with those delicate interactions, providing insightful comparison with $TaS_2$. For instance, monolayer 1T-$TaSe_2$ shows rich orbital textures [25] of its Mott state that are unavailable in $TaS_2$ [27]. For 1T-$NbSe_2$, the central Nb of its SD has stronger hybridization with its nearest Se than the case of $TaS_2$, resulting an amplification of the Mott gap via Jahn-Teller effect [28]. This implies the possibility of tuning its Mott gap via tailoring the crystal structure. As the Mott gap is governed by the ratio between hopping integral and Coulomb repulsion, the same parameter for determining the QSL regime [6],[7], its tuning may provide invaluable control knob for unraveling the exotic properties of such states of matter.

In this work, with spectroscopic imaging scanning tunneling spectroscopy (SI-STM), we characterize the Mott insulating state in monolayer 1T-$NbSe_2$ and realize its Mott gap tuning through structural distortion of the SD motif. The SDs exhibit temperature dependent switching with concomitant change in Mott gap size. The switching, as unveiled from first principles calculations, originates from atom coordination adjustment of the SDs. This causes variation of the wavefunction overlap between the central Nb orbital and the valence bands and hence generates different effective Coulomb energy. In addition, Mott collapse occurs in bilayer 1T-$NbSe_2$, as caused by interlayer coupling.

The experiments were performed with a custom made cryogenic Unisoku STM

system [29]. High quality monolayer and bilayer 1T-NbSe$_2$ films are grown by MBE on graphene-covered SiC(0001) substrate [Fig. S1]. The theoretical calculation is carried out with density functional theory (DFT) plus $U$. Detailed descriptions of the experiments and the calculations are depicted in Supplementary Information.

The 1T-NbSe$_2$ consists of a triangular lattice of Nb layer sandwiched between two Se layers, with each Nb atom octahedrally coordinated by six Se atoms. At low temperature, 1T-NbSe$_2$ enters a commensurate CDW phase, forming a $\sqrt{13} \times \sqrt{13}$ superlattice composed of SD motifs [Fig. 1a], similar to 1T-TaS$_2$. Each SD is centered with a Nb atom with six nearest and six outmost Nb atoms. Figs. 1b-c show the topography of the monolayer 1T-NbSe$_2$ film taken at 77 K. Homogeneous SD throughout the film can be seen under an imaging bias of -1 V. However, under +1V, STM image of the same film clearly displays two different types of SDs, whose apparent heights in the center differ by approximately 92 pm. Consequently, they are named as bright and dark SDs, respectively. Majority SDs are of bright type, and the dark SDs are mostly distributed around the periphery of the islands. Zoom-in STM image of the film clearly resolves the atomic resolution of the top layer Se atoms that conforms to the SD structure [Fig. S1c]. The two types of SDs have no discernable difference in their atomic resolution. Interestingly, when the film is cooled to 4.4 K, the bright SDs imaged at 77 K under +1 V drastically change to dark SDs [Fig. 1e]. In contrast, STM image of the SDs under -1 V keeps the same as that of the 77 K [Fig. 1d]. This indicates the $\sqrt{13} \times \sqrt{13}$ CDW state of the SDs are still reserved, suggesting the possibility of an electronic transition associated with the Mott state.

Such hypothesis is examined from the differential conductance spectra, which are proportional to the local density of states (LDOS) of the sample. For the two types of SDs imaged at 77 K, their spectra taken at the SD center exhibit distinct spectroscopic features [Fig. 1f]. The dark SD features a valence band (VB) peak at -280 meV and a band gap of 450 meV, where the VB maximum and the conduction band (CB) minimum locate at -30 meV and 420 meV, respectively. Above the CB edge, its conductance increases monotonously. Its band gap has been ascribed as a Mott gap in a previous study [26], whose specific spectral shape is, however, not identical to ours. For the bright SD, its VB peak shifts to -230 meV, but the CB features are similar to those of the dark SD. Importantly, there appears a pronounced low-energy peak located at 160 meV within the gap. This peak is responsible for the higher apparent height of the bright SDs at positive imaging bias. In accordance to its STM topography, the SD at 4.4 K has no peak at 160 meV [Figs. 1e, f], which resembles the dark SD at 77 K.

To understand the origin of the peak at 160 meV, we first performed spectroscopic mapping to the SDs. At 77 K, the d$I$/d$V$ conductance mapping of the bright SDs [Fig. 2a] at -230 meV (750 meV), corresponding to the VB (CB), displays highest (lowest) intensity at the central SDs [Figs. 2b, d]. Such conductance reversal conforms to the anti-phase relation of conductance at the two edges of the CDW gap [30]. The highest conductance intensity of the peak at 160 meV also locates at central SDs [Fig. 2c]. For the case of 1T-TaS$_2$, the Hubbard bands are contributed by the central Ta of the SD, which results the brightness in the SD center of its STM image [27],[31]. Hence, the peak at 160 meV here is ascribed to the upper Hubbard band (UHB). The shift of the UHB induces the 'dark'-'bright' SD transition, as is substantiated by our first principles calculations later. At 4.4 K, the spectroscopic mapping of the SDs [Fig. 2e] show similar features of as CB and VB conductance distributions as those at 77 K [Figs. 2f, h]. The spatial conductance intensity of the SD centers becomes triangular shaped at 4.4 K, in contrast to that of the 77 K, which is rounded, possibly due to the increased thermal broadening of the states. Notably, spectroscopic mapping of 160 meV at 4.4 K unveils a weak $(1/\sqrt{3} \times 1/\sqrt{3})R30°$ superstructure relative to that of the SDs [Fig. 2g], which is likely related to the precursor of the UHB (Supplementary Note 1).

The distinct behavior of the SDs at different temperatures leads us to track the temperature evolution of their electronic states. Figs. 3a-b show the temperature dependent conductance plots taken along a line traversing through the same monolayer SDs [Fig. S2a, black line]. As the temperature increases above 46 K, the SDs jump suddenly from the dark type into the bright type, despite a few SDs remain in the dark state [Fig. S3]. Concomitant switching occurs to the peak at 160 meV, i.e., the UHB [Fig. S2b]. At 2.2 K, the conductance plot displays spatial variations associated with the CDW state for both the CB and the VB [Fig. 3a]. At the SD centers, there is no peak at 160 meV, conforming to Fig. 1f. However, between the SDs, a small peak emerges at 160 meV, which is ascribed as the UHB precursor (Supplementary Note 1) and conforms to the spectroscopic mapping in Fig. 2g. Upon the SDs switching to the bright type, the enhanced peak at 160 meV emerges right at the SD centers [Fig. 3b], demonstrating the Mott gap concurrently decreases. To depict such transition quantitatively, we extracted the spectra of the SD centers from the 2D conductance plot at each temperature [Fig. 3c], and subsequently obtained the conductance intensities of the peak at 160 meV with a Lorentz fitting to its line shape [Fig. 3d]. The peak intensity changes within a narrow temperature range of < 10 K and keeps constant at both low and high temperatures.

To understand the observed abrupt Mott-gap transition, we carried out DFT calculations. Based on the total energy calculations, we obtained the most stable structure of the SDs in 1T-NbSe$_2$, whose phonon spectrum is shown in Fig. S4a. It corresponds to the 'dark' SD state at low temperature. Its calculated electron-phonon coupling indicates two strong low energy breathing modes of the SDs at 2.68 and 3.67 THz (Fig. 4a), which are on a similar energy scale of the measured switching temperature in experiment. Thus, we suggest that the electron-phonon coupling trigger a structural distortion abruptly with rising temperature, inducing the 'dark'-'bright' SD transition. Considering the phonon displacement pattern of the two breathing modes (Fig. S4b-c), we obtain a possible metastable structure of the SDs, corresponding to the 'bright' SD state at high temperature. Compared to the dark SD, the bright SD has lattice distortions, where the most influential coordinate adjustments on the Mott state involve the nearest Nb and Se. The nearest Nb atoms shrink towards the central Nb in plane, while the nearest Se atoms have both an in-plane shrinkage to the SD center and an out-of-plane expansion [Fig. 4b]. This structural distortion is counter intuitive to the Peierls distortion for CDW formation, which predicts an in-plane expansion of the nearest Nb atoms with increasing temperature. This demonstrates the electron-electron interaction outweighs the electron-phonon interaction in stabilizing the SD structure at low temperatures.

Based on the two different SD structures, we calculate their band structures, along with the simulated STM images [Figs. 4c-h]. For the dark SD structure, its UHB lies above the CB minimum [Fig. 4d], and its simulated STM image displays dark SDs [Fig. 4e]. In contrast, for the bright SD structure, its UHB is down-shifted deeply inside the bandgap [Fig. 4g], and its simulated STM image shows bright SDs [Fig. 4h], confirming to our experimental observations and the UHB-shift scenario. Therefore, our simulation demonstrates that the Mott gap transition of the SDs is indeed associated with their lattice distortions.

In the following, we scrutinize the mechanism of the Hubbard band shift induced by the lattice distortion. Figs. 4c,f indicate the dispersion of the localized band, which arises from the $dz^2$ orbital of the central Nb atom, in the two SD structures. Such localized $dz^2$ band is buried beneath the VB, reflecting hybridization between the Nb $dz^2$ orbital and the Se $p$ orbital. While the energy of the localized band relative to the VB maximum remains nearly unchanged at the two SD structures, its degree of localization exhibits marked difference. Namely, the localized band is nearly dispersionless in the dark SD [Fig. 4c], but significantly increases its bandwidth in the bright SD [Fig. 4f]. This is because the lattice distortion in the bright SD increases the hybridization between the Se-$p$ and Nb-$d$ orbitals. To quantitatively depict such hybridization, the wavefunction of the localized $dz^2$ band can be written as $|LB_i\rangle = \omega_{di}|d_i\rangle + \omega_{pi}|p_i\rangle$, $\omega_{di}$ ($\omega_{pi}$) is the wavefunction weight of the $d$ ($p$) orbital at the $i$-th SD. The effective Coulomb energy is given as $U_{eff} \sim \omega_{di}^4 \cdot U$ [32], whose derivation is described in Supplementary Note 2. The increasing width of the localized band decreases its $U_{eff}$, resulting in the down-shift of the UHB [Figs. 4d and g]. It is noted that the Hubbard band shift is also observed in a previous system 1T-TaS$_{2-x}$Se$_x$ [31], which mechanism is distinct to this case. In TaS$_{2-x}$Se$_x$, the Hubbard band shift is caused by a change of on-site energy difference in the Ta-$dz^2$ orbital from the Se substitution.

The above analysis demonstrates that the Mott gap in 1T-NbSe$_2$ is intimately connected to the SD structural distortion. This suggests applying strain as a viable approach of tuning its Mott gap, which can lead to a similar lattice distortion to the 'dark'-'bright' transition. Fig. S5 shows the calculated band structures of 1T-NbSe$_2$ under biaxial strain. With increasing the strain from -2% to 2%, the localized band of the SD gradually decreases its bandwidth [Fig. S5, upper panel]. Correspondingly, the UHB shifts upward [Fig. S5, lower panel], and the $U_{eff}$ changes monotonically with the strain amplitude [Fig. 4i]. In contrast to 1T-NbSe$_2$, 1T-TaS$_2$ has weak hybridization between the central Ta-$d$ orbital and the S-$p$ orbital, whose Mott gap and the associated $U_{eff}$ is thus insensitive to the strain [Figs. 4i and S6].

Finally, we have also investigated the interlayer coupling effect on the Mott state of the system by measuring bilayer NbSe$_2$. As shown in Fig. 5a, the second layer show the same SD pattern as the first layer. Zoom-in image resolves the outmost Nb of the second layer stacks on top of the central Nb atom of the first layer [Fig. 5b]. Similar stacking configuration has been observed in bulk TaS$_2$ [21,33] and bilayer TaSe$_2$ [25]. Interestingly, the Mott gap observed at the 1$^{st}$ layer NbSe$_2$ vanishes completely on the $2^{nd}$ layer, leaving a spectral dip around the Fermi level [Fig. 5c]. This suggests the interlayer coupling could increase the bandwidth of the localized $dz^2$ orbital, reducing the $U_{eff}$ and driving the system into a metal [21,[],33].

In summary, by combining SI-STM measurements and first principles calculations, we have investigated the Mott insulating state hosted in the lattice of SD motifs in monolayer 1T-NbSe$_2$, which demonstrates Mott collapse in the presence of interlayer coupling. The Mott gap exhibits intriguingly reversible switching, which is driven by a temperature induced structural distortion. Due to the stronger overlap between the central Nb $d$ orbital and the Se $p$ orbital, the structural distortion causes prominent changing of the $U_{eff}$ and the associated energies of Hubbard bands. This suggests the Mott gap of 1T-NbSe$_2$ can be tuned with strain, thereafter providing a viable approach for realizing QSL state. It can be surmised that similar effect can exist in related compounds possessing strong $p$-$d$ orbital coupling, such as monolayer 1T-TaSe$_2$; hence, offering a general strategy for tuning the Mott state via controlled structural distortions.

**Acknowledgement:** We thank Zheng Liu, Tetsuo Hanaguri and Bin Shao for discussions. This work was funded by the National Key Research and Development Program of China (Grant Nos. 2017YFA0403501, 2016YFA0401003 and 2018YFA0307000), the National Natural Science Foundation of China (Grant Nos. 11874161, U20A6002, 11774105, 12047508, 11634003) and NSAF under Grant No. U1930402.

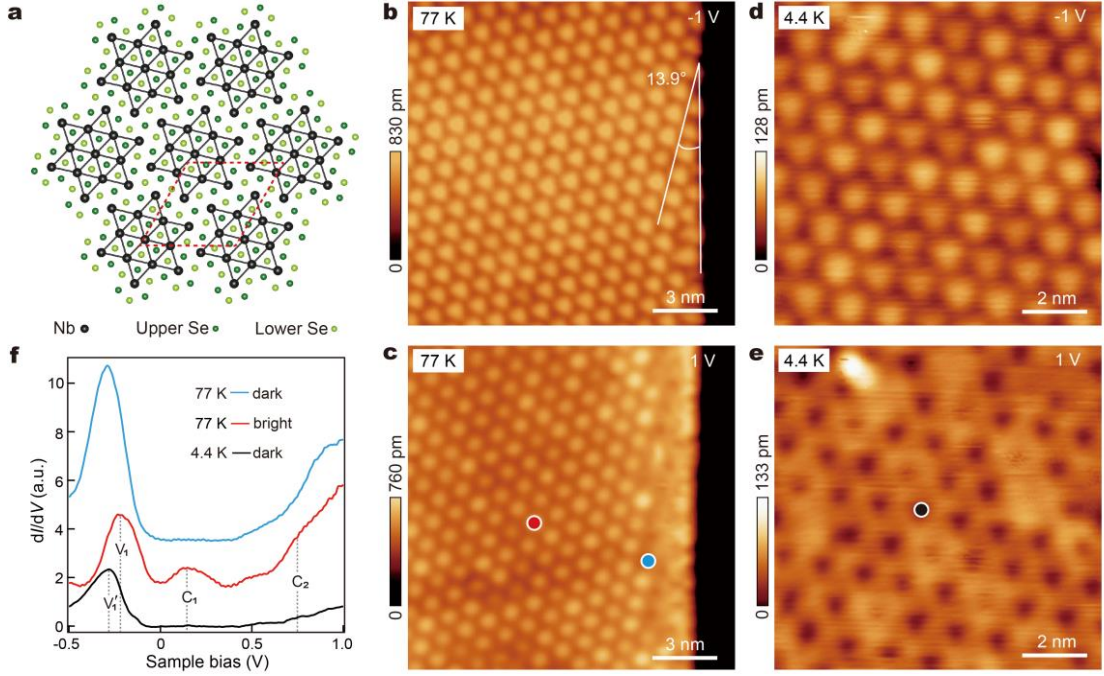

**Fig. 1 Two types of SDs. a** Structural sketch of monolayer 1T-NbSe$_2$. The SD motifs and CDW supercell are outlined by black and red lines, respectively. **b,c** STM images of the same monolayer 1T-NbSe$_2$ island obtained at 77 K with negative and positive imaging biases, respectively. Imaging conditions: **b** $V_b$ = -1.0 V and $I_t$ = 20 pA; **c** $V_b$ = +1.0 V and $I_t$ = 20 pA. **d,e** STM image of the same monolayer 1T-NbSe$_2$ island obtained at 4.4 K with negative and positive imaging biases, respectively. Imaging conditions: **d** $V_b$ = -1.0 V and $I_t$ = 20 pA; **e** $V_b$ = +1.0 V and $I_t$ = 20 pA. **f** d$I$/d$V$ spectra ($V_b$ = +1.0 V, $I_t$ = 200 pA) taken at the SD centers marked in **c** and **e**.

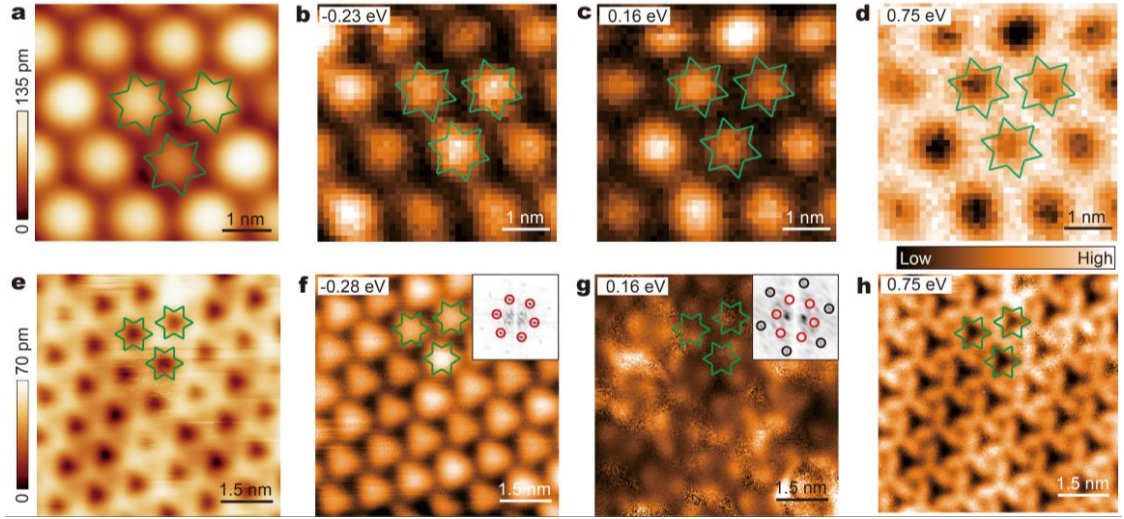

**Fig. 2 Conductance mappings at 77 K and 4.4K. a,e** STM image ($V_b$ = +1.0 V and $I_t$ = 20 pA) of a monolayer 1T-NbSe$_2$ obtained at 77 K and 4.4 K, respectively. **b-d** Spectroscopic conductance mappings of **a** at different energies.. **f-h** Constant height conductance maps of **e** at different energies. Insets of **f** and **g** show their respective fast Fourier transformation images. Red circles and black circles represent the diffraction spots of the SD lattice and the superstructure of the 160 meV state, respectively. Green hexagrams in all images mark typical SDs

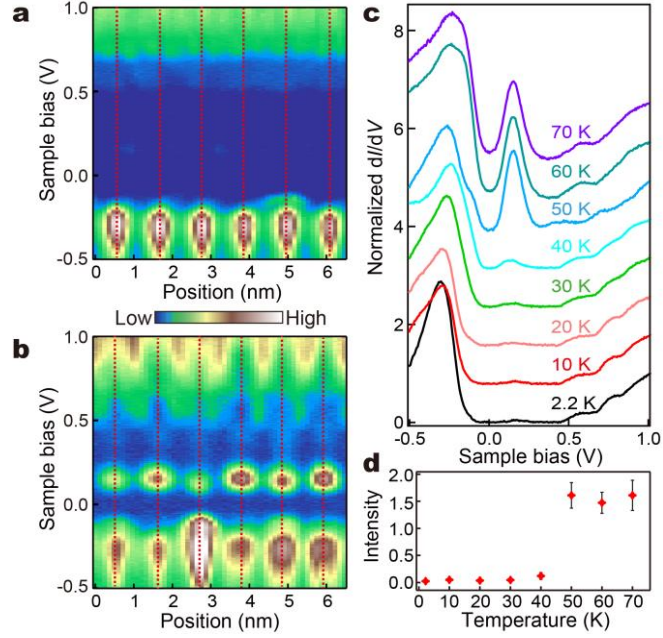

**Fig. 3 Temperature driven Mott gap transition. a,b** 2D conductance plot ($V_b$ = +1.0 V, $I_t$ = 50 pA) traversing through the same six SD centers at 2.2 K (**a**) and 50 K (**b**), respectively. The red dashed lines mark the SD centers. **c** Normalized spectra averaged from those of the SD centers, as exemplified in **a** and **b** at different temperatures. **d** Temperature dependent conductance intensity of the in-gap state extracted from **c**.

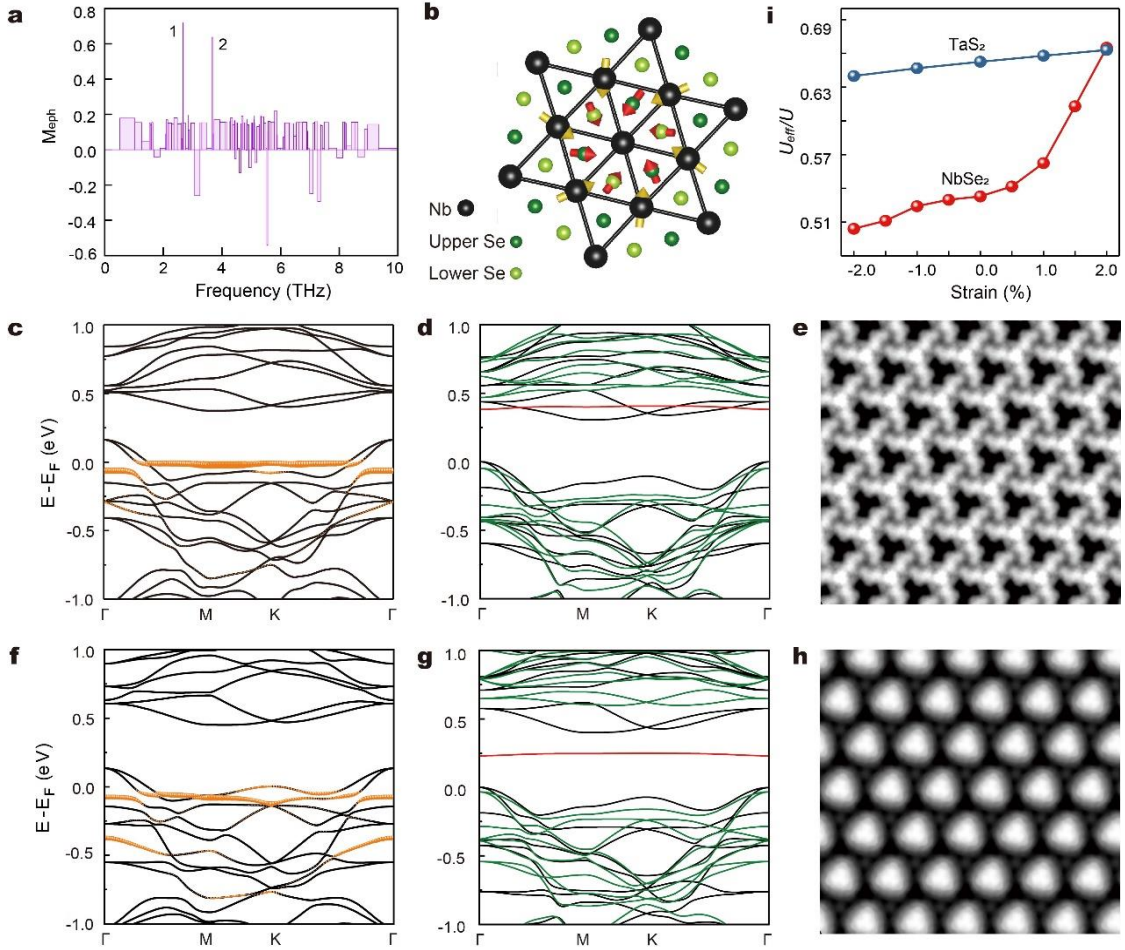

**FIG. 4 Calculated bands and simulated STM images. a** Electron-phonon coupling constant ($M_{eph}$) for different phonon modes at Γ. **b** Top view of the SD structure, showing the lattice distortion of nearest Nb (orange arrows) and Se (red arrows) atoms from the dark to bright SD transition. The arrow length depicts the distortion magnitude. **c,d** Band structure of the *dark* SDs for the nonmagnetic states (**c**) and spin polarized states (**d**), respectively. **f,g** Band structure of the *bright* SDs for the nonmagnetic states (**f**) and spin polarized states (**g**), respectively. In **c,f**, the localized $d_{z^2}$ band is highlighted with orange color. In **d,g**, the spin-up and spin-down are in green and black color, respectively. The UHB (spin-up) is highlighted with red color. **e,h** Simulated STM

image for the dark SDs (**e**) and bright SDs (**h**) based on the calculated bands of **d** and **g**, respectively. **i** Calculated $U_{eff}$ as a function of strain for monolayer 1T-NbSe$_2$ and monolayer 1T-TaS$_2$.

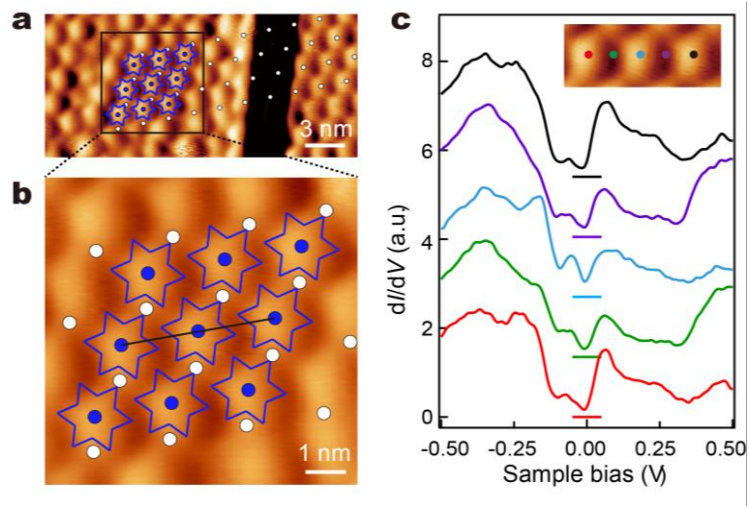

**Fig. 5 Mott collapse in bilayer film. a** STM image ($V_b$ = -1.0 V, $I_t$ = 50 pA) in differential mode showing the stacking configuration of bilayer NbSe$_2$. The white (blue) dots mark the SD centers of the first (second) layer. The blue hexagrams depict the SD structure of the 2$^{nd}$ layer. **b** Zoom-in view of the rectangle area in **a**. **c** d$I$/d$V$ spectra ($V_b$ = -0.5 V, $I_t$ = 100 pA, $T$ = 77 K) taken at the positions indicated in the inset image, which shows three SDs in **b** marked with a line segment.